\documentclass[%
 reprint, twocolumn,
 superscriptaddress,
 amsmath,amssymb,
 aps,
 prb,
]{revtex4-2}

\usepackage{graphicx}
\usepackage{dcolumn}
\usepackage{bm}
\usepackage{mhchem}
\usepackage{placeins}
\usepackage{siunitx}
\usepackage{layouts}
\usepackage{color}
\usepackage{makecell}

\begin{document}

\preprint{APS/123-QED}

\title{\textbf{Polarization-Dependent Raman Selection Rules in Sb$_2$S$_3$ from First Principles and Experiment} }

\author{Tobias Dierke}
\affiliation
{Department of Physics, Chair of Experimental Physics, Friedrich-Alexander-Universit\"at Erlangen-N\"urnberg (FAU), Staudtstr. 7, 91058 Erlangen, Germany}

\author{Michael H\"uttenkofer}
\email{michael.huettenkofer@fau.de}
\affiliation
{Department of Physics, Chair of Experimental Physics, Friedrich-Alexander-Universit\"at Erlangen-N\"urnberg (FAU), Staudtstr. 7, 91058 Erlangen, Germany}

\author{Stefan Wolff}
\affiliation
{Department of Physics, Chair of Experimental Physics, Friedrich-Alexander-Universit\"at Erlangen-N\"urnberg (FAU), Staudtstr. 7, 91058 Erlangen, Germany}

\author{Mingjian Wu}
\affiliation
{Institute of Micro- and Nanostructure Research \& Center for Nanoanalysis and Electron Microscopy (CENEM) Friedrich-Alexander-Universität Erlangen-Nürnberg, IZNF, Cauerstraße 3, 91058 Erlangen, Germany}

\author{Julien Bachmann}
\affiliation
{Chemistry of Thin Film Materials, Materials Chemistry Section, Department Chemistry and Pharmacy, Friedrich-Alexander-Universität Erlangen-Nürnberg, IZNF, Cauerstraße 3, 91058 Erlangen, Germany}

\author{Erdmann Spiecker}
\affiliation
{Institute of Micro- and Nanostructure Research \& Center for Nanoanalysis and Electron Microscopy (CENEM) Friedrich-Alexander-Universität Erlangen-Nürnberg, IZNF, Cauerstraße 3, 91058 Erlangen, Germany}

\author{Janina Maultzsch}
\email{janina.maultzsch@fau.de}
\affiliation
{Department of Physics, Chair of Experimental Physics, Friedrich-Alexander-Universit\"at Erlangen-N\"urnberg (FAU), Staudtstr. 7, 91058 Erlangen, Germany}

\date{\today}

\begin{abstract}
Antimony sulfide (Sb$_2$S$_3$) is a semiconductor composed of quasi-one-dimensional ribbon-like structural units, which give rise to pronounced structural anisotropy in the bulk crystal. Despite growing interest in Sb$_2$S$_3$, in particular Sb$_2$S$_3$ thin films, a detailed understanding of its symmetry-based lattice dynamics  remains incomplete. 
Here, we present a combined experimental and theoretical study of polarization-dependent Raman scattering in Sb$_2$S$_3$ thin films. We derive the Raman selection rules from the crystal symmetry and calculate the zone-center phonon modes and corresponding Raman tensors using density functional theory. The calculated polarization dependencies are systematically compared with polarization-dependent Raman measurements performed on oriented crystalline domains of Sb$_2$S$_3$ thin films. This combined analysis enables reliable mode assignments, elucidates the anisotropic Raman response associated with the ribbon-like crystal structure, and demonstrates the sensitivity of polarized Raman spectroscopy to crystal orientation and structural order in antimony chalcogenide (Sb$_2$S$_3$, Sb$_2$Se$_3$) as well as isostructural Bi$_2$S$_3$ thin films.

\end{abstract}

\maketitle

\section{\label{sec:introduction}Introduction}

Antimony sulfide (\ce{Sb2S3}) is a fascinating low-dimensional van der Waals (vdW) semiconductor of high technological importance, as it is used in applications such as thin-film photovoltaic devices \cite{Parize2017, Bttner2019,Pawar2022, Hamburger2025}, photodetectors \cite{Deng2023,Shen2023}, and energy storage systems \cite{Kong2016, Dong2017,Nithya2024}. It can be synthesized using various methods, including chemical vapor deposition, spray pyrolysis, and hydrothermal evaporation \cite{Eensalu2019,Eensalu2022,Pawar2022,Shen2023,Abraham2023}. 
\ce{Sb2S3} is typically found as a bulk material consisting of quasi-one-dimensional, ribbon-like units \cite{Barthwal2022,Tobi2025}. These ribbons are oriented in parallel and are held together by relatively weak vdW interactions. Due to this weak inter-ribbon coupling, \ce{Sb2S3} exhibits pronounced anisotropy in its structural, electronic, and optical properties \cite{Li2021}. The anisotropic character of \ce{Sb2S3} is already apparent in polarized-light optical microscopy images of thin-film surfaces, where domains with different crystallographic orientations are visible.

A straight-forward method to determine the actual crystallographic orientation is based on probing the anisotropic vibrational properties via polarization-dependent Raman spectroscopy~\cite{sereni10, Fleck2020-su, Tobi2025}. Raman spectroscopy has been widely employed to characterize \ce{Sb2S3}, {\it e.g.}, in studies on crystal growth~\cite{Parize2017-ba}, coherent phonon dynamics~\cite{Chong2014-jk}, and phase transitions~\cite{Delaney2020-ga}.  However, despite several Raman investigations of \ce{Sb2S3} reported in the literature, inconsistent assignments of the  Raman-active modes have been reported \cite{Sorb2015,Ibz2016,MedinaMontes2016, Parize2017,Bttner2022, Gilshtein2025}. These discrepancies hinder the clear interpretation of experimental data and impede a comprehensive understanding of the material’s lattice dynamics, including the determination of crystallographic orientation, strain analysis, or thermal properties. Therefore, achieving a reliable and unambiguous assignment of the Raman modes is essential.

In this work, we present angular-resolved polarized Raman spectroscopy experiments on \ce{Sb2S3} thin films with precisely characterized crystallographic orientations, prepared by atomic layer deposition (ALD), together with density-functional theory (DFT) calculations. The comparison between experiment and theory enables an unambiguous assignment of the Raman modes observed. Additionally, 360\(^\circ\) polar plots show that the intensities measured for the Raman-active \ce{Sb2S3} modes agree with the theoretical values calculated using the Raman tensors of the individual modes. Based on these polar plots of the \ce{Sb2S3} modes, we can accurately assign the vibrational symmetry to the individual modes and thereby provide the basis for future investigations of the fundamental properties of \ce{Sb2S3} crystals and thin films, as well as isostructural materials like Sb$_2$Se$_3$ and Bi$_2$S$_3$ or the orthorhombic phase of Bi$_2$Se$_3$~\cite{Mediavilla-Martinez2025-ep}.

\section{\label{sec:setup}Fundamentals and Experimental Setup}

\begin{figure*}
    \centering
    \includegraphics[width=0.99\linewidth]{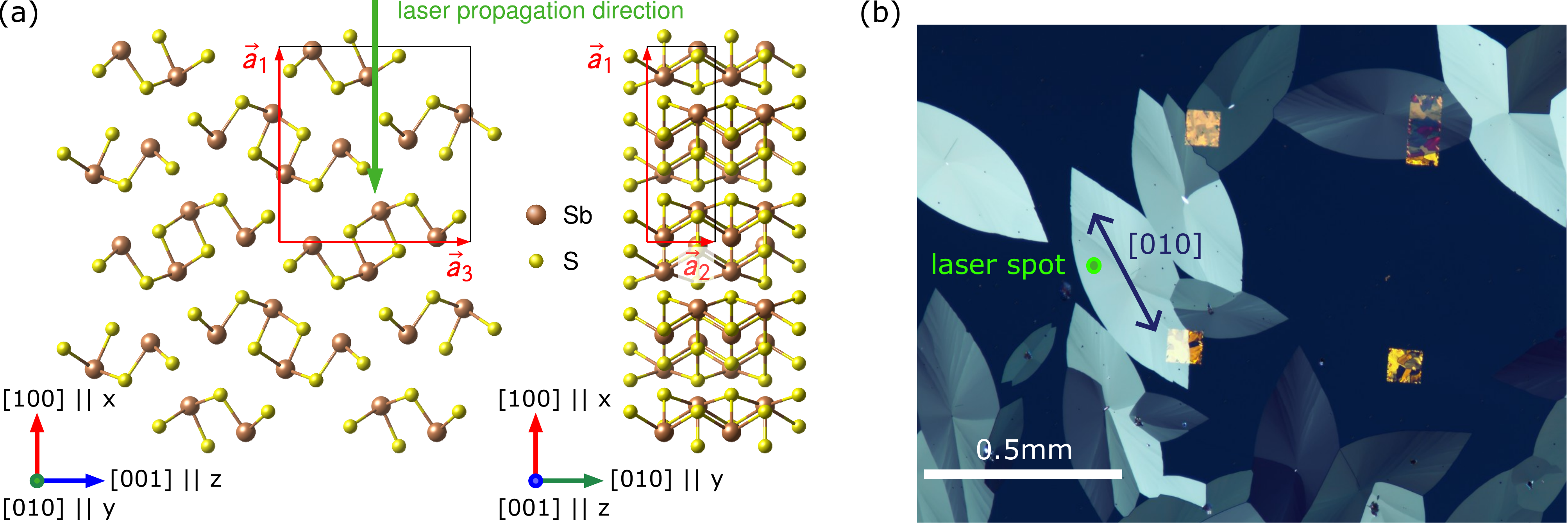}
    \caption{(a) Crystal structure of \ce{Sb2S3} projected onto the [010] and [001] crystallographic directions. These correspond to the $xz$ and $xy$ planes in the lab frame. The  coordinate system is the same as in Ref.~\cite{Tobi2025}, such that the ribbons extend along the [010] direction. The unit cell and lattice vectors $\mathbf{a}_1$, $\mathbf{a}_2$, and $\mathbf{a}_3$ are highlighted. For the Raman measurements, the laser propagates along the crystallographic [100] direction  and is therefore polarized within the $yz$ plane in the lab frame; a polarization angle of $0^\circ$ corresponds to a relative alignment of the polarization vector along the crystallographic [010] direction ($y$ axis), i.e., a polarization of the electric field along the ribbons. (b) Linear polarized light microscopy image of the samples for Raman study. The crystallographic directions of the “leaves” were  examined by SEM-EBSD analysis showing that the [010] direction aligns with the long axis of the “leaves” and the [100] direction with the surface normal (cf. Ref~\cite{Tobi2025}). The green dot indicates the location of the laser spot in all Raman measurements.}
    \label{fig:structure}
\end{figure*}

\subsection{Crystal Structure of \ce{Sb2S3}}

Antimony sulfide (naturally occurring as the mineral stibnite) crystallizes in the orthorhombic \textit{Pnma} space group (no. 62; $D_{2h}$ point group), giving rise to quasi-one-dimensional $[\ce{Sb4S6}]_n$ ribbons along the crystallographic [010] direction (see Fig.~\ref{fig:structure}). Within these ribbons, the atoms are covalently bonded, whereas adjacent ribbons are coupled via vdW interactions. Figure~\ref{fig:structure} (a) shows the crystal structure of \ce{Sb2S3}. The unit cell contains 20 atoms (8 antimony and 12 sulfur). The lattice vectors $\mathbf{a}_1$, $\mathbf{a}_2$, and $\mathbf{a}_3$, oriented along the $x$, $y$, and $z$ directions (lab frame), were obtained from DFT structural relaxations performed with and without vdW corrections. The inclusion of vdW corrections has a stronger impact on the lattice vectors $\mathbf{a}_1$ and $\mathbf{a}_3$ (decrease by 4.4\% and 2.9\%, respectively) than on $\mathbf{a}_2$ (decrease by 1.6\%), consistent with the expectation that vdW interactions are more significant along the $\mathbf{a}_1$ and $\mathbf{a}_3$ directions. See Figure~\ref{fig:basis} in appendix~\ref{sec:crystal_basis} for the lattice vectors and atomic coordinates obtained from DFT. The crystallographic directions [uvw] and the laboratory coordinate system $(xyz)$ shown in Fig.~\ref{fig:structure} are chosen to be consistent with Ref.~\cite{Tobi2025} and are used throughout this work. Note that in general, the crystallographic directions [uvw] do not coincide with the Cartesian  axes $x$, $y$, and $z$. However, due to the orthorhombic crystal structure of \ce{Sb2S3}, the crystallographic axes are mutually orthogonal and can therefore be chosen to coincide with the Cartesian coordinates used here. Defining this laboratory frame is particularly important for the description of light polarization vectors in the Raman intensity calculations (see Sec.~\ref{sec:geometry}).

The 20 atoms in the unit cell give rise to 60 phonon modes for each wave vector in the Brillouin zone. The $D_{2h}$ point-group symmetry results in four Raman-active representations: $A_g,\, B_{1g},\, B_{2g},\,\mathrm{and}\,\, B_{3g}$. The full vibrational representation at the $\Gamma$ point decomposes into the following irreducible representations:
\begin{equation}
\begin{split}
\Gamma ={}& 10\,A_g + 5\,A_u + 5\,B_{1g} + 10\,B_{1u} \\
& + 10\,B_{2g} + 5\,B_{2u} + 5\,B_{3g} + 10\,B_{3u}
\end{split}
\end{equation}
Therefore, in total there are 30 Raman active ($A_g,\, B_{1g},\, B_{2g},\, B_{3g}$), five silent ($A_u$), and 25 infrared-active modes ($B_{1u},\, B_{2u},\, B_{3u}$), including the three acoustic modes ($B_{1u},\, B_{2u},\, B_{3u}$).
Note that if the convention of the \textit{Pbnm} space group is used, the ribbon direction is along [001] and B$_{1g}$ and B$_{2g}$ are interchanged.

\subsection{Geometry of the Raman Measurement}\label{sec:geometry}

The \ce{Sb2S3} samples used in this work were thin-film \ce{Sb2S3} single crystallites of hundreds of micrometers in lateral size with [100] out-of-plane crystallographic direction. For further details about  preparation, see methods. 

\begin{figure}
    \centering
    \includegraphics[width=0.99\linewidth]{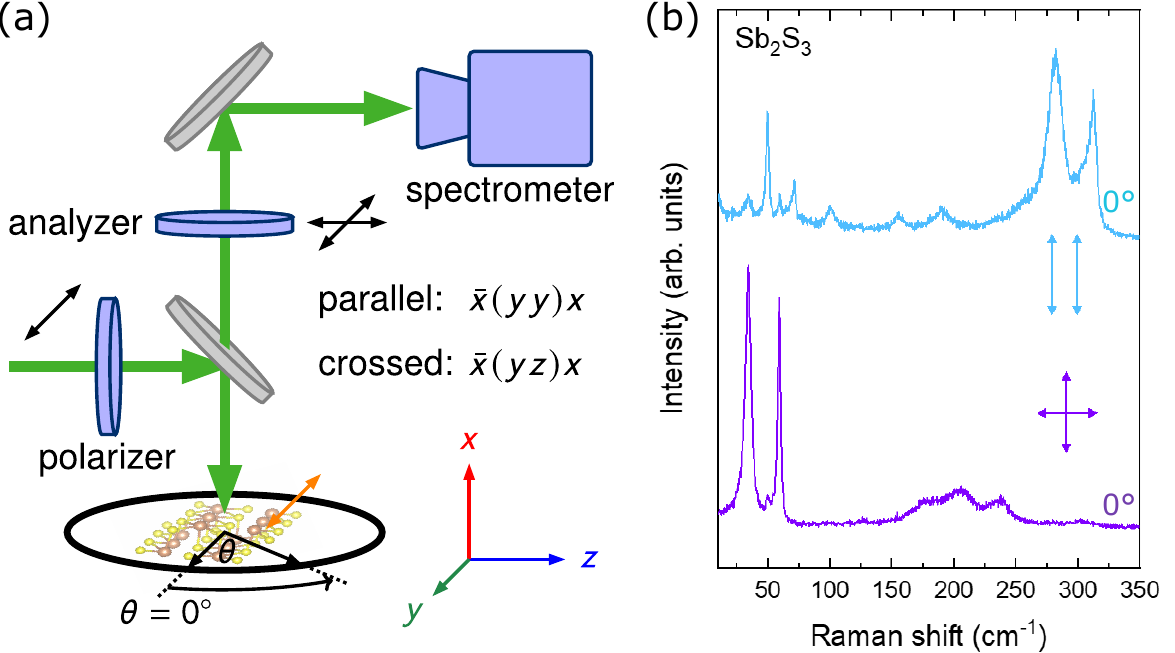}
    \caption{(a) Experimental setup of the polarization dependent Raman measurement. (b) Raman spectra for a polarization angle of $\theta=0^\circ$ in  parallel and crossed configuration}
    \label{fig:setup}
\end{figure}

All Raman measurements are performed in backscattering geometry with control of the polarization of incident and scattered light [see Fig. \ref{fig:setup}(a)]. The linear polarization of the incident laser beam is kept fixed, while the scattered light passes through an analyzer oriented either parallel or perpendicular (crossed) to the polarization of the incident beam. The sample is mounted on a rotational stage and can be rotated about the propagation direction of the incident beam. In this way, the crystallographic axes are rotated with respect to the fixed polarization of the incoming and scattered light, with the angle $\theta$ describing this relative orientation. This configuration allows for determining the symmetry of the observed Raman modes as shown in Sec. \ref{sec:discussion}.

In the experiments presented here, the incident laser beam propagates along the $x$ axis (crystallographic [100] direction) of the crystal. Consequently, the electric field of the laser light is polarized within the [0vw] plane of the crystal ($yz$ in the lab frame). The corresponding polarization vectors for the incident ($i$) and scattered ($s$) light in parallel ($\parallel$) and crossed ($\perp$) configurations are then given by:
\begin{equation}\label{equ:polvectors}
    \mathbf{e}_i=\left(\begin{array}{c}
            0\\
            \cos(\theta)\\
            \sin(\theta)
        \end{array}\right), \mathbf{e}_s^{\parallel}=\left(\begin{array}{c}
            0\\
            \cos(\theta)\\
            \sin(\theta)
        \end{array}\right), \mathbf{e}_s^{\perp}=\left(\begin{array}{c}
            0\\
            -\sin(\theta)\\
            \hphantom{-}\cos(\theta)
        \end{array}\right)
\end{equation}
We align our samples such that $\theta = 0^\circ$ corresponds to an incident laser polarization parallel to the [010] axis of the crystal. For $\theta = 0^\circ$, the parallel and crossed configurations in Porto's notation read $\bar{x}(y,y)x$ and $\bar{x}(y,z)x$, respectively.

For our Raman scattering geometry with the excitation laser propagating along the $x$ direction, only the $A_g$ and $B_{3g}$ modes yield non-zero Raman intensities. The corresponding Raman tensors for these symmetries take the general form \cite{Aroyo2006}.

\begin{equation}
    A_g=\begin{pmatrix}
    a & 0 & 0 \\
    0 & b & 0 \\
    0 & 0 & c
    \end{pmatrix}
    \qquad
    B_{3g}=\begin{pmatrix}
    0 & 0 & 0 \\
    0 & 0 & f \\
    0 & f & 0
    \end{pmatrix}
\end{equation}

Based on the $\theta$-dependent polarization vectors, the Raman intensity of mode~$\nu$ is calculated (Eq. \ref{equ:intensity}) according to Ref.~\cite{YuCardona}, using the Raman tensor $R^{\nu}$ of mode $\nu$ extracted from the DFT Raman calculations:
\begin{equation}\label{equ:intensity}
    I_s^\nu \propto \left| \mathbf{e}_i \cdot R^\nu \cdot \mathbf{e}_s \right|^2.
\end{equation}
A more detailed explanation of the DFT-calculated Raman tensor is given in the appendix \ref{sec:raman_tensor_dft}.

\section{\label{sec:discussion}Discussion}

\begin{figure*}
    \centering
    \includegraphics[width=0.99\linewidth]{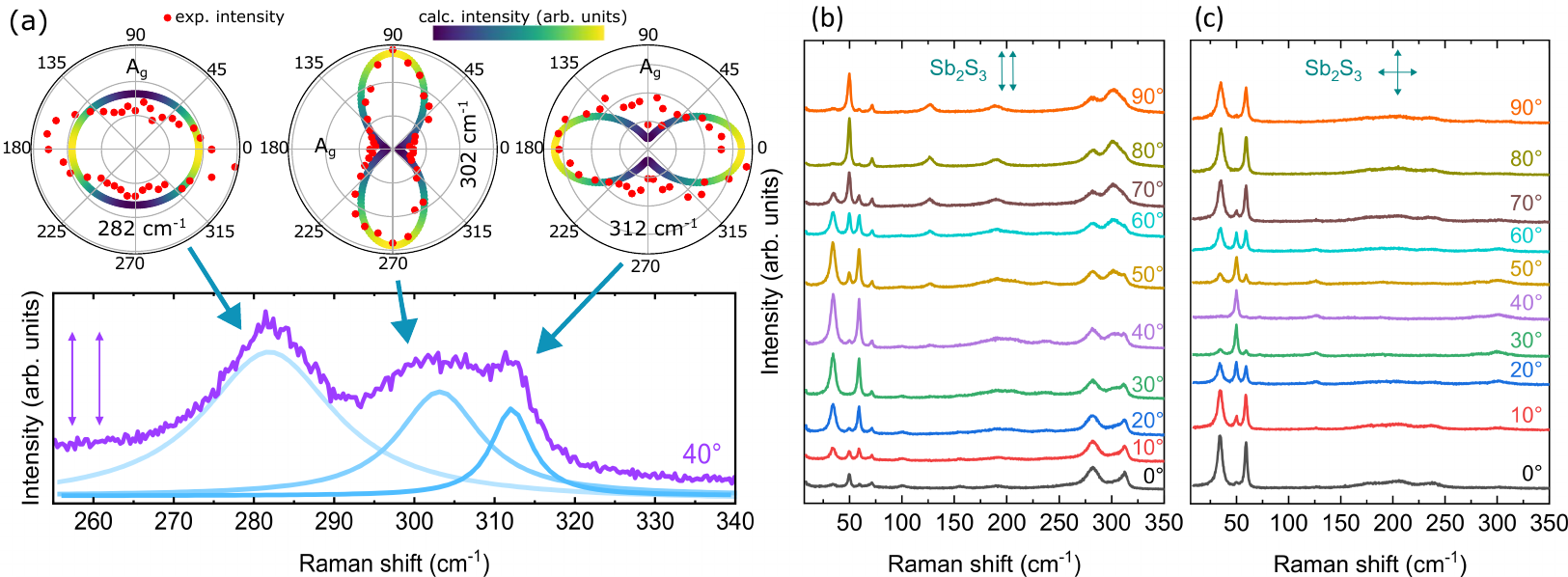}
    \caption{(a) Raman intensities of the three high-frequency $A_g$ modes (modes \#51, \#53, and \#56) at an experimental polarization angle of $\theta=40^\circ$. The symmetry assignment of these modes is based on polarization-dependent Raman measurements (polarization plots above the modes) and comparison with theoretical Raman tensor calculations obtained from DFT using Eqs.~\ref{equ:polvectors} and~\ref{equ:intensity}. 
    The intensity on the color scale ranges from the lowest calculated value (blue) to the largest calculated value (yellow) for modes \#51, \#53, and \#56, respectively. (b,c) Raman spectra of \ce{Sb2S3} measured under parallel (b) and crossed (c) polarization configurations for $\theta$ between $0^\circ$ and $90^\circ$.}
    \label{fig:raman_pol}
\end{figure*}

Figure~\ref{fig:setup}\,(b) displays the Raman spectra of the \ce{Sb2S3} film measured under an in-plane rotation angle of 0\(^\circ\) in parallel (purple) and crossed (blue) polarization configurations. The in-plane crystallographic orientation of the \ce{Sb2S3} crystallite film was determined before by EBSD (electron backscatter diffraction), as described in Fig.\,3 of Ref. \cite{Tobi2025}.

Figure~\ref{fig:raman_pol}\,(b) and~\ref{fig:raman_pol}\,(c) present Raman spectra from \(\theta=0^\circ\) to \(\theta=90^\circ\) in parallel and crossed polarization configuration (see Fig.~\ref{fig:structure}). 
The relative intensities of the Raman modes exhibit a pronounced dependence on the in-plane rotation angle, as expected from the anisotropy of the
\ce{Sb2S3} crystal.

To enable a consistent assignment of the Raman modes, the phonon frequencies (table~\ref{tab:frequencies}), atomic displacement patterns (see appendix~\ref{B2} Fig.~\ref{fig:raman_modes_2}), and Raman tensors of the \ce{Sb2S3} phonon modes are calculated by DFT, see~\ref{sec:comp_meths}. Details of the derivation of the Raman tensors from the DFT output are provided in the appendix~\ref{sec:raman_tensor_dft}. The calculated Raman tensors are subsequently used to simulate polarization-dependent Raman intensities according to Eq.~\ref{equ:intensity}. The predicted polarization-dependent  intensities are then compared with the experimentally determined Raman intensities, which are extracted from Lorentzian fits to the spectra. The experimental intensity is defined by the integrated peak area. Since the comparison focuses on the polarization dependence rather than on absolute intensity values, the computed intensities are scaled for each Raman mode by a single factor.

The three high-frequency modes at 282\,cm\(^{-1}\), 302\,cm\(^{-1}\), and 312\,cm\(^{-1}\) are usually the focus of studies on \ce{Sb2S3} due to their prominence. Based on our polarization-dependent Raman measurements, we assign all three modes as $A_g$, in our DFT calculations modes \#51, \#53 and \#56.
Figure~\ref{fig:raman_pol}\,(a) shows the three high-frequency \ce{Sb2S3} Raman modes measured in parallel  configuration for an in-plane rotation angle of 40\(^\circ\) in more detail. The polarization-dependent Raman intensities of these modes are presented as 360\(^\circ\) polar plots, together with corresponding simulations.
Our assignment to $A_g$ symmetry is further supported by the comparison between parallel and crossed polarization configurations, see Fig.~\ref{fig:raman_pol}\,(b,c), where the intensities of these modes in crossed polarization are strongly suppressed.

The assignment of the high-frequency Raman modes discussed above in \ce{Sb2S3} varies significantly across the literature. In particular, the modes at 282\,cm\(^{-1}\), 302\,cm\(^{-1}\), and 312\,cm\(^{-1}\) have in some reports been incorrectly assigned to $B_{3g}$ symmetries \cite{Bttner2022,Ibz2016,Sorb2015, Gilshtein2025,MedinaMontes2016}. In contrast to this, the combined experimental and DFT-based polarization analysis presented here enables for an unambiguous and consistent symmetry assignment of these modes as $A_g$.
Furthermore, as previously shown in Ref.~\cite{Tobi2025}, the in-plane crystallographic orientation of \ce{Sb2S3} can be determined from the relative intensity ratios of these high-frequency Raman modes (\#51, \#53, \#56). This simplifies the analysis of the crystal orientation of \ce{Sb2S3} crystals from single Raman spectra (see appendix~\ref{B1} Fig.~\ref{fig:intensityratio}).

\begin{figure*}
   \centering
   \includegraphics[width=0.95\linewidth]{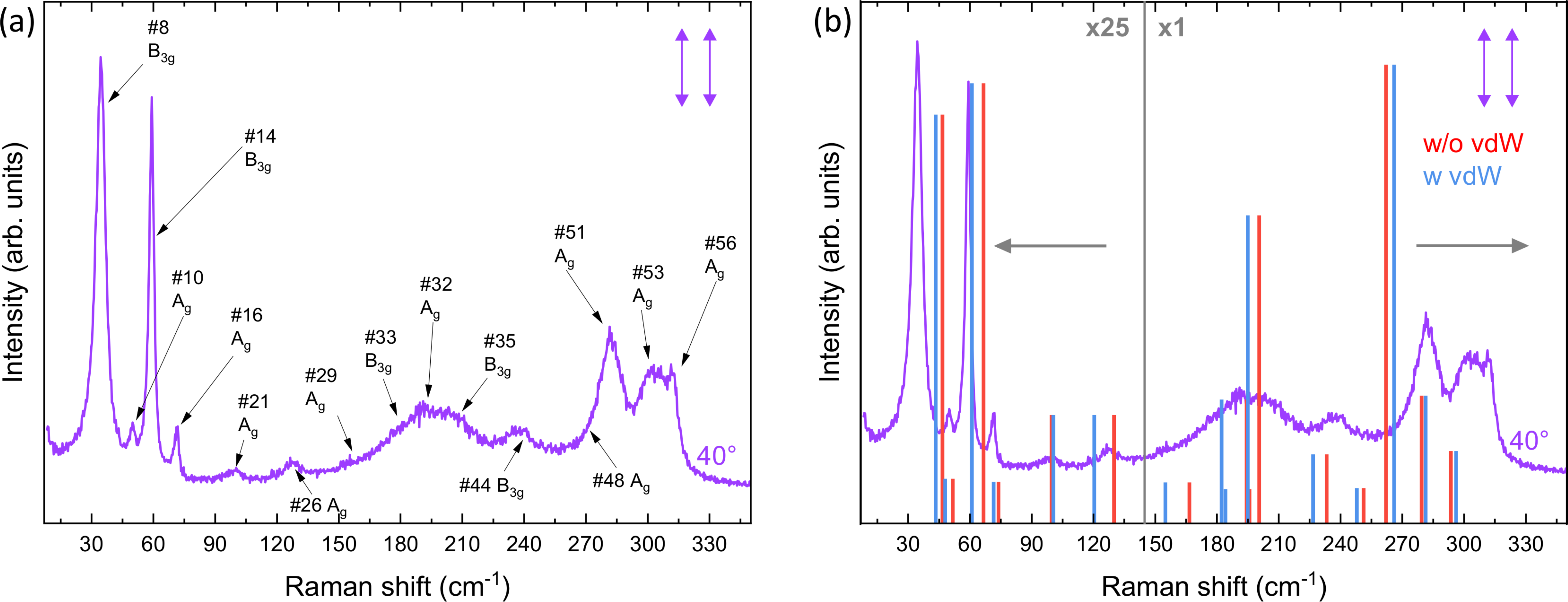}
   \caption{(a) Experimental Raman spectrum measured in parallel polarization at a polarization angle of 40$^\circ$, with symmetry assignments based on calculated phonon frequencies and polarization-angle dependent Raman intensities obtained from DFT. (b) Experimental Raman spectrum (purple) measured in parallel polarization geometry at a polarization angle of $40^\circ$, compared with calculated Raman intensities and phonon frequncies (red: without vdW corrections, blue: with vdW corrections) evaluated at $47.5^\circ$ using Eq.~\ref{equ:intensity}. To improve visibility, the calculated intensities in the left part of the spectrum are multiplied by a factor of 25 relative to those on the right-hand side.} 
   \label{fig:intensities_exp_theo}
\end{figure*}

The polarization plots of all experimentally accessible Raman modes, measured in parallel configuration, are provided in appendix~\ref{B1}, Fig.~\ref{fig:parallelpolplots}.
In general, the experimental data presented in the polar plots are in good agreement with the theoretical predictions.
Fig.~\ref{fig:intensities_exp_theo} shows the Raman spectrum of \ce{Sb2S3} at an in-plane rotation angle of 40\(^\circ\), where all Raman-active modes are observed; their symmetries are assigned based on our polarization analysis.

A detailed examination of the polarization behavior of the two $A_g$ modes at 71\,cm\(^{-1}(\#16)\) and 100\,cm\(^{-1}(\#21)\) indicates that the entries of their Raman tensors are interchanged by the DFT calculations, see polar plots appendix~\ref{B1} Figs.~\ref{fig:parallelpolplots} and ~\ref{fig:32+35}\,(a). Furthermore, small deviations between experiment and theory are observed for the $A_g$ mode at 190\,cm\(^{-1}(\#32)\) and the neighboring $B_{3g}$ mode at 206\,cm\(^{-1}(\#35)\), arising from their close spectral proximity, which complicates simultaneous fitting. By summing up the experimental intensities from both fits as well as the calculated intensities, a better agreement between experiment and theory is achieved [see appendix~\ref{B1} Fig.~\ref{fig:32+35}\,(b)].

The crossed-polarized Raman data are shown in appendix~\ref{B1} Fig.~\ref{fig:crosspolplots}. These measurements further corroborate the assignment of the $B_{3g}$ modes, while the $A_g$ modes are strongly suppressed and only weakly detectable for most in-plane orientations. 
In fact, the intensities of the $A_g$ modes vanish at rotation angles of 0\(^\circ\) and 90\(^\circ\) in the crossed-polarized configuration [see Fig.~\ref{fig:setup} (a)] and are very weak for all other angles [see Fig.~\ref{fig:raman_pol}(c)]

Small discrepancies between the experimental and calculated polarization dependencies are observed for a few Raman modes. These deviations can occur for modes that are spectrally close and therefore require simultaneous fitting. In such cases, the extracted intensities of a weak mode can be sensitively influenced by that of a neighboring, much stronger mode. This happens for example for modes \#51 and \#56 at 90$^\circ$ and 270$^\circ$, where mode \#53 exhibits a particularly strong Raman response, while modes \#51 and \#56 are comparatively weak. As a result, their fitted intensities may be affected by the spectral overlap with the tail of mode \#53. 

\noindent
\begin{table*}
    \caption{Calculated frequencies of the Raman-active phonon modes of \ce{Sb2S3} compared with experimental Raman measurements. The mode number corresponds to its index within the complete set of 60 phonon modes. Values given in brackets denote the mode index obtained from calculations including vdW interactions, reflecting the reordering of phonon modes due to changes in the calculated phonon frequencies. Experimental peak positions are determined from Lorentzian fits to the Raman spectra. All frequencies are given in units of cm$^{-1}$.}
    \label{tab:frequencies}
    \begin{ruledtabular}
    \begin{tabular}{c c c c c }
            mode number & \makecell{frequency DFT\\without vdW}  & \makecell{frequency DFT\\with vdW} & experimental frequency &symmetry  \\\hline
            8 (9) & 47 & 44 & 34 & B$_{3g}$ \\
            10 & 52 & 48 & 50 & A$_g$ \\
            14 & 67 & 61 & 59 & B$_{3g}$ \\
            16 (17) & 74 & 72 & 71 & A$_g$ \\
            21 & 100 & 101 & 100 & A$_g$ \\
            26 (25) & 130 & 121 & 128 & A$_g$ \\
            29 (30) & 167 & 155 & 158 & A$_g$ \\
            32 & 195 & 182 & 190 & A$_g$ \\
            33 & 196 & 184 & 177 & B$_{3g}$ \\
            35 & 201 & 195 & 206 & B$_{3g}$ \\
            44 & 233 & 227 & 238 & B$_{3g}$ \\
            48 (47) & 251 & 248 & 258 & A$_g$ \\
            51 & 262 & 266 & 282 & A$_g$ \\
            53 & 280 & 282 & 303 & A$_g$ \\
            56 & 293 & 296 & 312 & A$_g$ \\
    \end{tabular}
    \end{ruledtabular}
\end{table*}

To assess the level of agreement between experiment and theory, Raman intensities of the Raman-active modes were calculated for a parallel polarization geometry at a polarization angle of $47.5^\circ$ and directly compared with the experimental Raman spectrum measured at nominally $\theta = 40^\circ$ in parallel polarization, as shown in Fig.~\ref{fig:intensities_exp_theo}. The theoretical intensities are evaluated at $47.5^\circ$ to account for a systematic angular offset of approximately $5^\circ$–$10^\circ$ observed between the experimental and calculated polarization dependencies. This offset most likely originates from a slight misalignment of the crystal with respect to the intended initial crystallographic orientation during the experiment.

For several modes, both the positions of the calculated Raman peaks and their relative intensities agree reasonably well with the experimental data (see Fig.\ref{fig:intensities_exp_theo}). Noticeable deviations of the vibrational frequency occur only for the two prominent low-frequency modes with $B_{3g}$ symmetry (modes \#8 and \#14) and for the three high-frequency modes with $A_g$ symmetry (modes \#51, \#53 and \#56). For the low-frequency $B_{3g}$ modes, the calculated frequencies are higher than the experimentally determined frequencies, whereas the calculated values of the high-frequency $A_g$ modes are smaller than the experimental ones.
As a first hypothesis, these deviations in frequency are attributed to the absence of vdW interactions in the DFT calculations. To study this effect, we performed an additional calculation in which vdW interactions were included in the structural relaxation before calculating the phonon frequencies [without including the vdW interaction into the phonon calculations to reduce computational cost (see Sec.~\ref{sec:comp_meths})]. The resulting changes in the phonon frequencies are summarized in Table~\ref{tab:frequencies}.
Indeed, the vibrational frequencies of the $B_{3g}$ and $A_g$ modes discussed above shift in the expected direction (closer to the experimental values), although quantitative agreement with experiment is still not fully achieved (see Tab~\ref{tab:frequencies}). A full Raman-tensor calculation including vdW corrections is not possible within the DFT framework employed in this work. Consequently, only the phonon frequencies could be corrected, while the Raman intensities remain unchanged. However, we do not expect strong changes in the Raman intensities upon inclusion of vdW corrections. Overall the calculation shown in Fig.~\ref{fig:intensities_exp_theo} shows that vdW corrections are relevant for the structural parameters and they improve agreement between experimental and calculated phonon frequencies. A closer discussion of the effects of vdW interactions on the phonon calculations can be found in appendix~\ref{A2}.

DFT results and group theory predict 15 further Raman-active modes ($B_{1g}$ and $B_{2g}$ symmetry) in \ce{Sb2S3} that are not observable with the present backscattering configuration. Access to these modes would require a different excitation geometry, for example changing the propagation direction of the incident light along the $y$ or $z$ axis (see Fig~\ref{fig:structure}).
We should note that the Raman modes obtained from DFT calculations in previous studies (e.g., reference~\cite{Ibz2016}) are generally accurate. However, due to the close-by frequencies of several phonon modes, experimental peaks were often incorrectly assigned. In the present work, we resolved this ambiguity by polarization-dependent Raman measurements, which allow for a reliable distinction between $A_g$ and $B_{3g}$ modes with similar frequencies.

\section{\label{sec:Conclusion}Conclusion}
In summary, we present polarization-dependent Raman measurements on an ALD-grown \ce{Sb2S3} thin film together with DFT-based calculations of the polarization-dependent Raman intensities. The good agreement between experiment and theoretical predictions enables a complete assignment of all experimentally accessible \ce{Sb2S3} Raman modes. This work establishes a reliable framework for future Raman studies of \ce{Sb2S3} and related materials such as \ce{Sb2Se3} and \ce{Bi2S3}. It provides accurate mode identification and contributes to a comprehensive understanding of their vibrational properties and pronounced anisotropy. More broadly, it introduces a robust methodology to study Raman selection rules and assigning Raman modes to their correct symmetry by combining DFT-based Raman-tensor calculations with polarization-dependent Raman measurements.

\section{\label{sec:methods}Methods}

\subsection{Experimental Methods}
\textbf{Raman spectroscopy}\\
Raman spectroscopy of the \ce{Sb2S3} sample was performed using a HORIBA LabRam HR Evolution spectrometer with a laser wavelength of \(\lambda=532.17\)\,nm (2.33\,eV), laser focus \(<\)\SI{1}{\micro\metre}, an 1800 lines/mm and an 3000 lines/mm grating. To avoid heating effects, laser powers below 0.5\,mW were used. Polarization-dependent Raman experiments were performed using a polarizer and an analyzer placed in the optical path to control the incident and scattered light polarizations. The \ce{Sb2S3} sample was mounted on a rotation stage to vary its in-plane orientation. All spectra were calibrated by neon lines. 

\textbf{Atomic Layer Deposition of \ce{Sb2S3}}\\
\ce{Sb2S3} was deposited using a homemade hot-wall atomic layer deposition (ALD) reactor onto Si/SiO\(_x\) TEM grids with SiO\(_x\) windows, following established protocols \cite{Tobi2025,Bttner2022,Bttner2021}. The precursors used were tris-(dimethylamido)antimony(III) \ce{(Sb(NMe2)3}, 99.99\%, Sigma-Aldrich), and \ce{H2S} (3\% vol in \ce{N2}, Air Liquide). Nitrogen was used as the carrier gas, and the reaction temperatures were 120\,\(^{\circ}\)C. The precursors were kept at room temperature, and the opening, exposure, and pumping times were 1.5\,s, 15\,s, and 15\,s for \ce{(Sb(NMe2)3} (kept at 40\,\(^{\circ}\)C) and 0.2\,s, 15\,s, and 15\,s for \ce{H2S}. In experiments, 600 ALD cycles were used to deposit \ce{Sb2S3} (yielding \(\sim\)\,36\,nm film thickness on planar substrates as measured by spectroscopic ellipsometry).
The samples were amorphous phase as deposited, and subsequently annealed up to 250\,\(^{\circ}\)C in the vacuum of a transmission electron microscope. The amorphous thin film transformed to the crystalline phase, and the crystalline orientation of specific crystallites was systematically characterized using electron back-scattered diffraction (EBSD) in the scanning electron microscope (SEM) \cite{Tobi2025}. Prior to the Raman study in this work, the crystal orientation of identical crystallites was confirmed again by EBSD.

\subsection{Computational Methods}\label{sec:comp_meths}
Density functional theory (DFT) calculations were performed using the Quantum ESPRESSO suite~\cite{QE_2009,QE_2017}. The geometric optimization of the Sb$_2$S$_3$ structure was performed until the forces exerted on each atom were below 0.001\,eV\AA$^{-1}$. Converged parameters for the plane-wave cutoff of 80\,Ry (1088.46\,eV) and a $5\times15\times5$ Monkhorst-Pack $k$-point grid to sample the Brillouin zone were utilized. All calculations employed either the local density approximation (LDA) for the exchange–correlation functional (calculations without van der Waals (vdW) corrections) or the generalized gradient approximation (GGA, PBE) for calculations including vdW corrections. Norm-conserving pseudopotentials from the SPMS library were used throughout,~\cite{SPMS2023}. Each lattice vector was allowed to change freely during the geometric optimization. In order to estimate the effect of vdW interactions, the relaxation of the structures was calculated with and without the DFT-D3 vdW correction,~\cite{DFT-D3}.

A Density Functional Perturbation Theory (DFPT) was used to calculate the phonon frequencies, Raman intensities, and displacement patterns at the $\Gamma$ point of the Brillouin zone. The starting geometries are the relaxed structures with and without vdW interactions. The inclusion of vdW interactions for the relaxation of the structure significantly affects the phonon frequency, whereas the inclusion of vdW interactions within the DFPT calculations did not result in notable changes. Therefore we did not include vdW interactions in the calculation of phonon frequencies (except those shown in Fig.\ref{fig:intensities_exp_theo} and Table~\ref{tab:frequencies}), keeping the computational cost at a minimum.

\begin{acknowledgments}
This work was financially supported by Deutsche Forschungsgemeinschaft (German Research Foundation, DFG) within the Collaborative Research Center ‘ChemPrint’ (project 538767711, CRC 1719), the Research Training Group 'CorMic' (project 537140136,
GRK3103), and project 447264071 (INST 90/1183-1 FUGG). The authors gratefully acknowledge the scientific support and HPC resources provided by the Erlangen National High Performance Computing Center (NHR@FAU) of the Friedrich-Alexander-Universität Erlangen-Nürnberg (FAU) under the NHR project b181dc. NHR funding is provided by federal and Bavarian state authorities. NHR@FAU hardware is partially funded by the German Research Foundation (DFG) – 440719683.

Artificial intelligence (AI) tools (ChatGPT, version 4 and 5) were used to assist with technical questions regarding Fortran code.
\end{acknowledgments}

\appendix

\section{Additional Information}
\subsection{Raman Tensor from DFT}\label{sec:raman_tensor_dft}
In Quantum ESPRESSO (QE), the quantity labeled as the Raman tensor in \textit{lraman} calculations represents the derivative of the macroscopic dielectric susceptibility with respect to small atomic displacements (DFPT). Accordingly, the Raman tensor for a phonon mode~$\nu$, as defined in~\cite{YuCardona}, can be written as:
\begin{equation}\label{equ:raman_tensor}
R_{ij}^{\nu}=\sum_{k=1}^{N_A}\sum_{\alpha\in(x,y,z)}
\underbrace{\frac{\partial\chi_{ij}}{\partial u_{k,\alpha}}}_{\mathrm{QE\ output}}\frac{e^\nu_{k,\alpha}}{\sqrt{M_k}}
\end{equation}

where $\chi_{ij}$ is the electric susceptibility tensor, $u_{k,\alpha}$ the atomic displacement of atom $k$ in Cartesian direction $\alpha$, $e^\nu_{k,\alpha}$ the (not normalized) eigenvector component of phonon mode~$\nu$ for atom $k$ and Cartesian direction~$\alpha$, and $M_k$ is the mass of atom~$k$.

Since Quantum ESPRESSO internally employs the quantity defined in Eq.~\ref{equ:raman_tensor} only to compute Raman intensities within the Placzek approximation~\cite{Walter2019}, i.e., without explicit polarization dependence, we modified the source code to directly obtain as output the full Raman tensor $R_{ij}^{\nu}$. With the full tensor, Raman intensities are calculated using Eqs.~\ref{equ:polvectors} and~\ref{equ:intensity} and subsequently compared with polarization-dependent Raman measurements. 

\subsection{Effect of vdW interactions on the DFT calculations}\label{A2}

All Raman calculations presented in this work were performed without vdW corrections, since Quantum ESPRESSO does not support vdW interactions in the calculation of Raman tensors. Nevertheless, to study the influence of vdW interactions on the vibrational properties of \ce{Sb2S3}, additional calculations were carried out focusing on the phonon frequencies.
Two different approaches were employed. First, the lattice parameters were relaxed including DFT-D3 vdW interactions, while the subsequent phonon calculations were performed without vdW corrections. This approach allows the indirect effect of vdW interactions through the modified equilibrium structure to be evaluated. Second, using a more recent version of Quantum ESPRESSO (version 7.5), phonon frequencies were calculated directly, including DFT-D3 vdW corrections. 

The modes exhibiting the largest deviations between experiment and theory are the two low-frequency modes with $B_{3g}$ symmetry (modes \#8 and \#14) and the three high-frequency modes with $A_g$ symmetry (modes \#51, \#53 and \#56), as already discussed in the main part of the paper. A plausible origin of these discrepancies is the absence of van der Waals interactions in the Raman DFT calculations.

For modes with $B_{3g}$ symmetry, the atomic displacements occur along the crystallographic $y$ axis, i.e., along the $[\ce{Sb4S6}]_n$ ribbons. In this case, additional van der Waals interactions acting between neighboring ribbons along the $x$ and $z$ directions may effectively weaken the restoring forces associated with vibrations along the ribbon direction. As a consequence, neglecting vdW interactions leads to an overestimation of the corresponding force constants and therefore to a blueshift of the calculated phonon frequencies compared to experiment.

For modes with $A_g$ symmetry, the atomic displacements predominantly lie within the $xz$ plane. In this case, vdW interactions between adjacent ribbons are expected to increase the effective restoring forces associated with these displacements, which would lead to higher phonon frequencies and a redshift of the calculated frequencies when vdW interactions are neglected.
It should be noted, however, that the influence of vdW interactions is strongly mode dependent and difficult to quantify in detail. From a qualitative perspective, vibrations dominated by heavier atoms are expected to be more sensitive to vdW interactions than those dominated by lighter atoms. This provides a possible explanation for the fact that not all $A_g$ or $B_{3g}$ modes are affected to the same extent: low-frequency modes are typically dominated by antimony displacements, whereas high-frequency modes involve predominantly sulfur vibrations.

A quantitative analysis of how strongly individual phonon modes in \ce{Sb2S3} are affected by van der Waals interactions is challenging due to the structural complexity of the crystal. Nevertheless, phonon calculations that include vdW interactions clearly show that the five modes discussed above shift in the expected direction, thereby supporting the qualitative interpretation presented here [see Fig.~\ref{fig:intensities_exp_theo} (b) and Tab.~\ref{tab:frequencies}]. At the same time, the inclusion of vdW interactions also leads to shifts in other phonon modes that previously showed good agreement with experiment. This indicates that vdW corrections do not uniformly improve the agreement for all modes, but maybe also introduce new complications within the calculation of phonon frequencies.

\section{Figures}
\subsection{Polar plots of polarization-angle dependent Raman intensities}\label{B1}
\begin{figure*}
    \centering
    \includegraphics[width=0.92\linewidth]{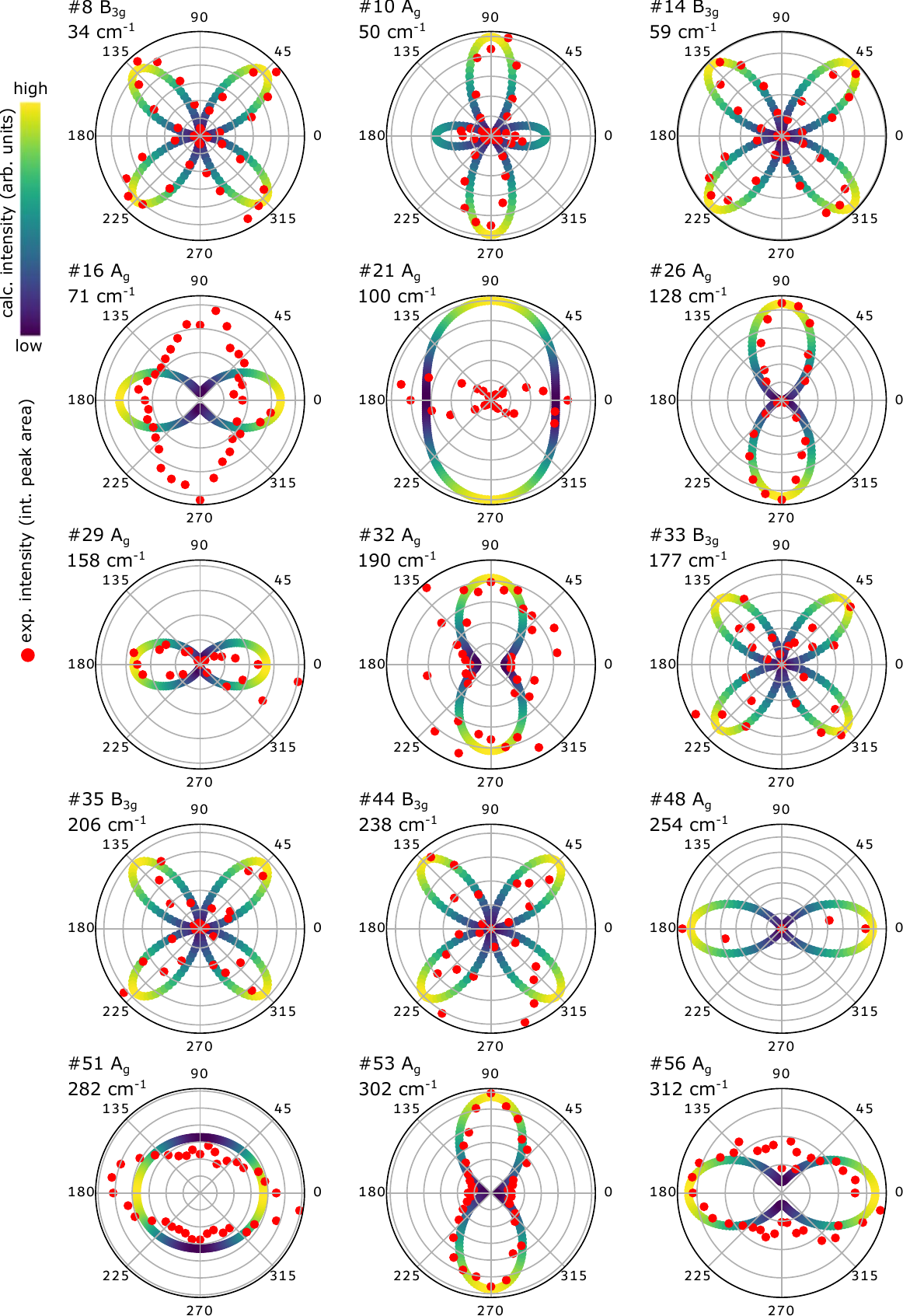}
    \caption{Polar plots in \textit{parallel} polarization configuration for all Raman active modes.}
    \label{fig:parallelpolplots}
\end{figure*}

\begin{figure*}
    \centering
    \includegraphics[width=0.92\linewidth]{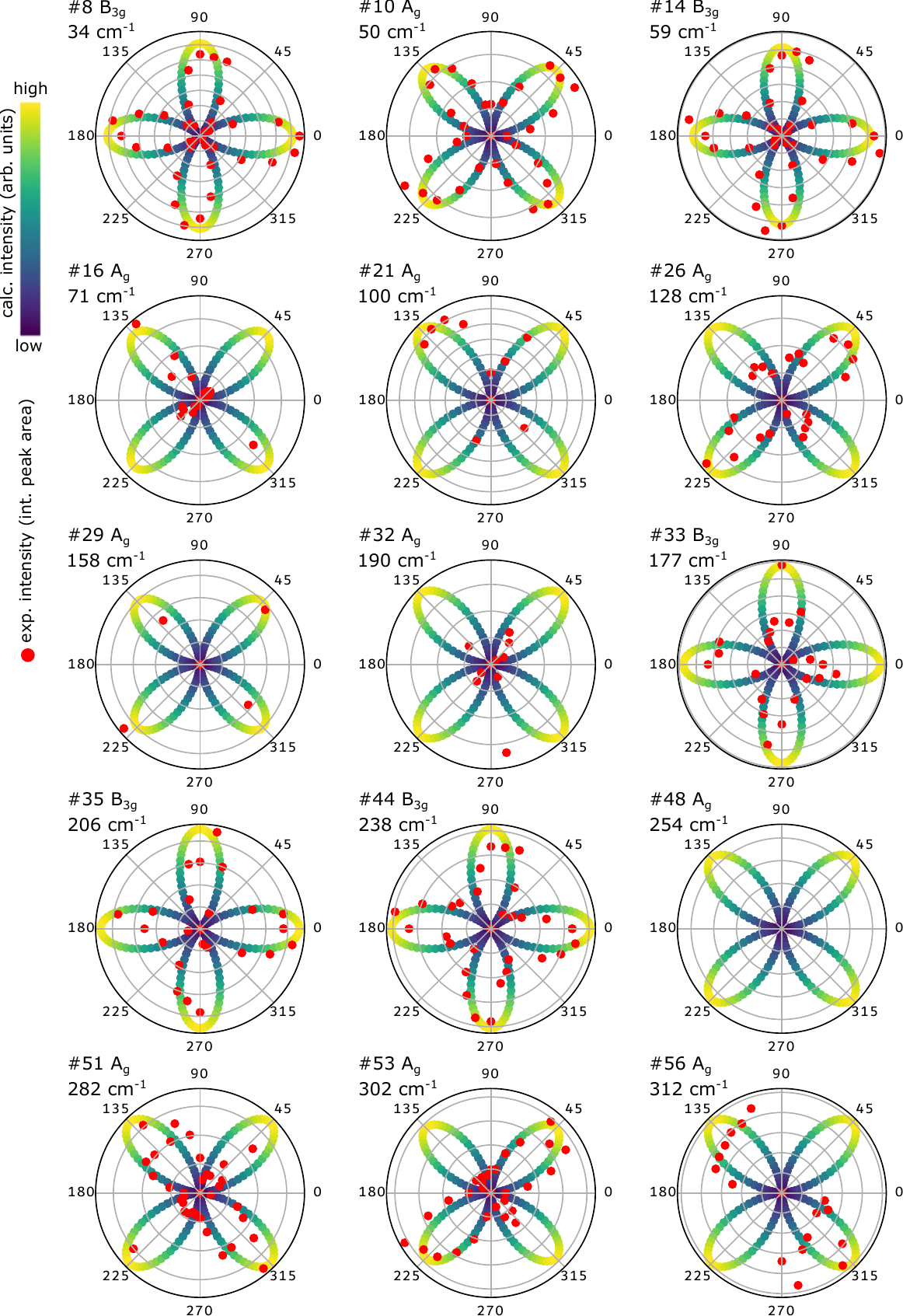}
    \caption{Polar plots in \textit{crossed} polarization configuration for all Raman active modes.}
    \label{fig:crosspolplots}
\end{figure*}

\begin{figure*}
    \includegraphics[width=0.95\textwidth]{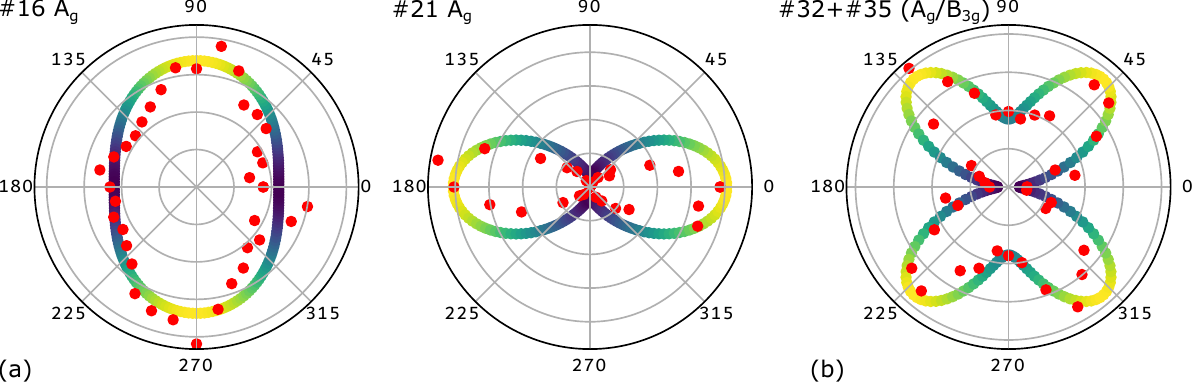}
    \caption{(a) Exchange of DFT models between modes \#16 and \#21 in parallel configuration. The simulations now clearly match the experimental data. (b) Combined presentation of modes \#32 and \#35 in parallel configuration. The experimental intensities of both modes are summed and compared with the corresponding summed DFT-calculated intensities. The combined theoretical intensity is scaled by a single factor.}
    \label{fig:32+35}
\end{figure*}

\begin{figure*}
    \centering
    \includegraphics[width=0.8\linewidth]{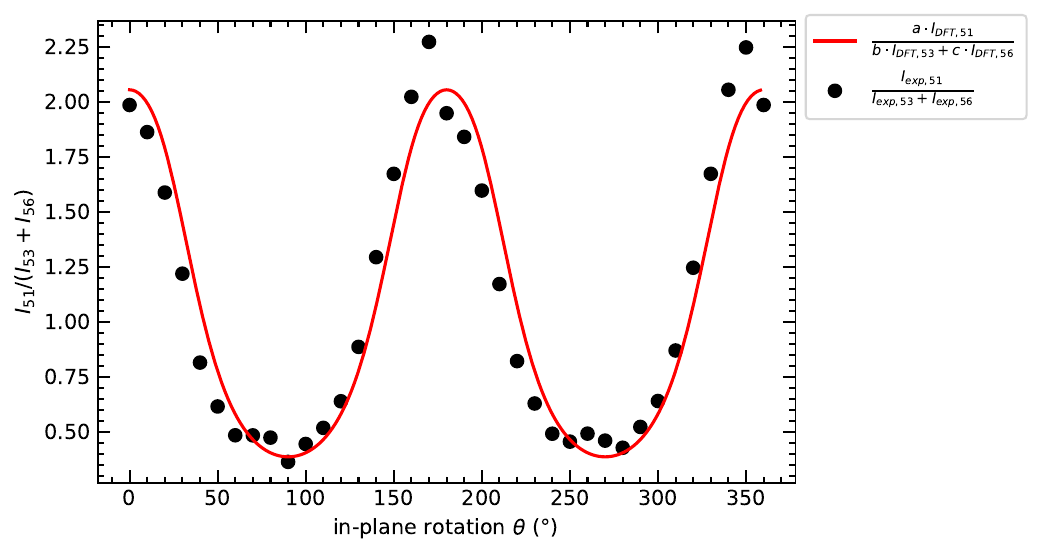}
   \caption{Intensity ratio of the high-frequency Raman modes in parallel polarization configuration as a function of the in-plane orientation of the \ce{Sb2S3} crystal. The figure shows the intensity of mode \#51 normalized to the combined intensity of modes \#53 and \#56. The red curve represents the DFT-calculated intensity ratio, fitted to the experimental data (black dots) using the expression given in the legend. The resulting fit parameters are $a=2.804$, $b=12.919$, and $c=4.598$.}
    \label{fig:intensityratio}
\end{figure*}

Here we show the experimental and theoretical polarization-dependent Raman intensities for all Raman active modes of \ce{Sb2S3} as polar plots in parallel (Fig.~\ref{fig:parallelpolplots}) and crossed (Fig.~\ref{fig:crosspolplots}) polarization configuration. In addition we also show an exchange of DFT models between modes \#16 and \#21 in parallel configuration [Fig.~\ref{fig:32+35} (a)], a combined presentation of modes \#32 and \#35 in parallel configuration [Fig.~\ref{fig:32+35} (b)], as well as the intensity ratio values (parallel configuration) of the high-frequency modes in relation to the changing in-plane orientation of the \ce{Sb2S3} crystal (Fig.~\ref{fig:intensityratio}).\\
All theoretical polarization-dependent Raman intensities for all 60 phonon modes at the $\Gamma$-point can be found in the Supplementary Information. This includes also the calculated Raman tensors for each mode.

\subsection{Displacement Patterns of Raman active phonon modes}\label{B2}
\begin{figure*}
    \centering
    \includegraphics[width=0.95\linewidth]{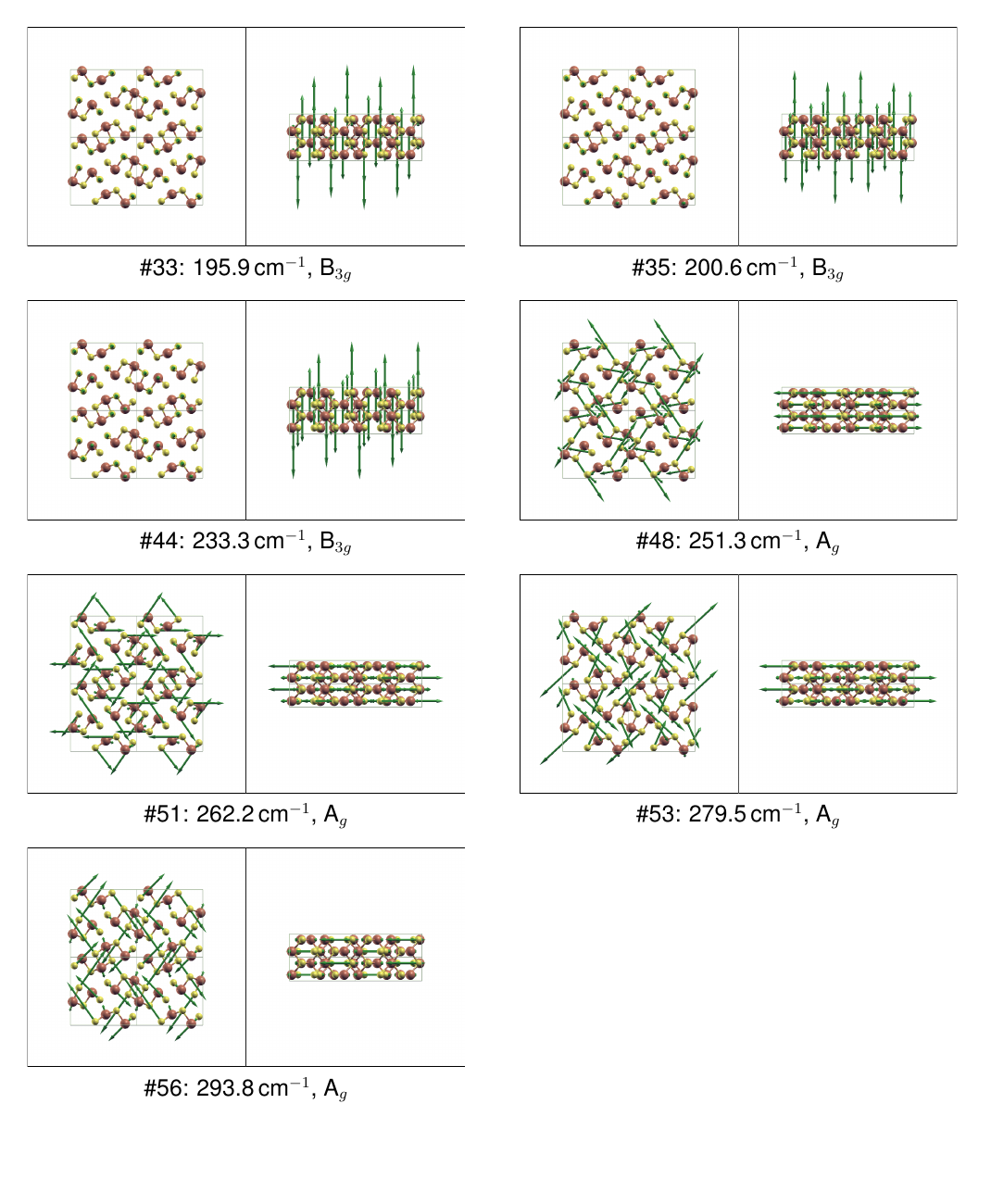}
    \caption{Displacement patterns of all Raman-active phonon modes of \ce{Sb2S3} in $xz$-plane and $yz$-plane.}
    \label{fig:raman_modes_2}
\end{figure*}

Here we show the displacement patterns of the phonon modes of \ce{Sb2S3} for all Raman-active modes in $xz$-plane and $yz$-plane with their corresponding phonon frequency and Raman symmetry (Fig.~\ref{fig:raman_modes_2}). The displacement patterns, phonon frequencies, and Raman symmetries have been calculated in a DFT phonon calculation without vdW corrections and been visualized in xcrysden~\cite{xcrysden}. The displacement patterns of all modes are shown in the SI
\begin{figure*}[t]
    \centering
    \includegraphics[width=0.95\linewidth]{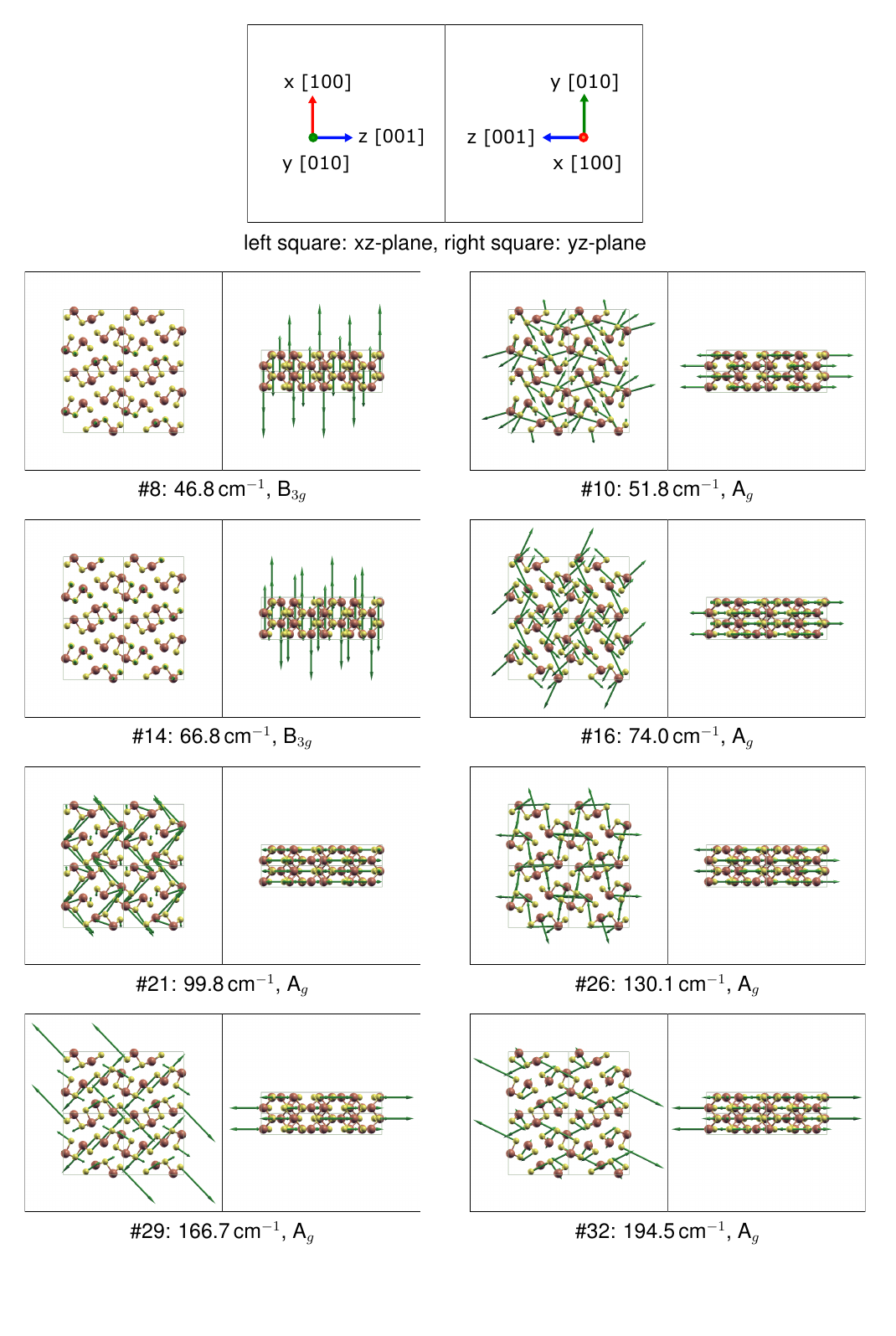}
    \label{fig:raman_modes_1}
\end{figure*}

\subsection{Crystal Basis of \ce{Sb2S3}}\label{sec:crystal_basis}
\begin{figure*}
    \centering
    \includegraphics[width=0.48\linewidth]{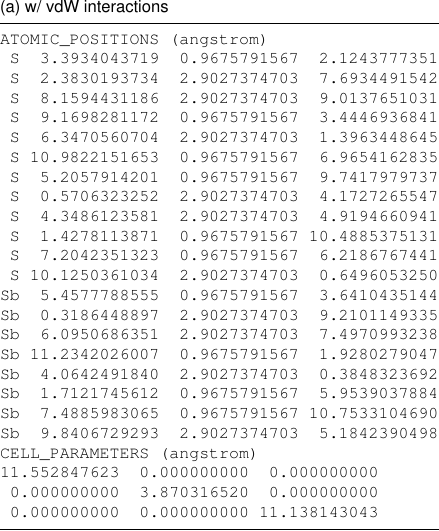}
    \hfill
    \includegraphics[width=0.48\linewidth]{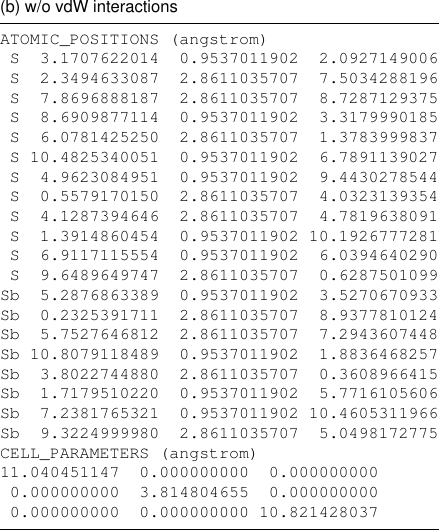}
    \caption{Atomic coordinates $(x,y,z)$ of the crystal structure of \ce{Sb2S3} obtained from DFT relaxations including (a) and excluding (b) vdW corrections. Length units are \AA.}
    \label{fig:basis}
\end{figure*}

Here we show the relaxed atomic coordinates and lattice vectors in Cartesian coordinates ($x,y,z$) of the crystal structure of \ce{Sb2S3}, which have been used for the DFT phonon and Raman calculations (Fig.~\ref{fig:basis}). The coordinates have been relaxed with and without vdW corrections. The length units are in Angström [\AA].

\FloatBarrier
\section{References}
\bibliography{bibliography}

\end{document}